\documentclass[runningheads]{llncs}
\usepackage{graphicx}
\usepackage{booktabs}
\usepackage{multirow}
\usepackage{diagbox}
\usepackage{caption}
\usepackage{subcaption}

\begin{document}
\title{Penalizing small errors using an Adaptive Logarithmic Loss}
%
%
\author{Chaitanya Kaul\inst{1}\orcidID{0000-0003-4893-6222}
 \thanks{Chaitanya Kaul and Roderick Murray-Smith acknowledge support from the iCAIRD project, funded by Innovate UK (project number 104690).} \and
Nick Pears\inst{2}\orcidID{0000-0001-9513-5634} \and
Hang Dai\inst{3}\orcidID{0000-0002-7609-0124} \and
Roderick Murray-Smith\inst{1}\orcidID{0000-0003-4228-7962} \and
Suresh Manandhar\inst{4}\orcidID{0000-0002-2822-2903}}
\authorrunning{C. Kaul et al.}
%
\institute{School of Computing Science, University of Glasgow, United Kingdom, G12 8RZ \\
\email{\{Chaitanya.Kaul,Roderick.Murray-Smith\}@glasgow.ac.uk}\\ \and
Department of Computer Science, University of York, United Kingdom, YO10 5DD \\
\email{nick.pears@york.ac.uk}\\ \and
MBZUAI, Masdar City, Abu Dhabi, United Arab Emirates\\
\email{hang.dai@mbzuai.ac.ae} \\ \and
NAAMII, Katunje, Bhaktapur, Kathmandu, Nepal\\
\email{suresh.manandhar@naamii.org.np}} 

\maketitle              
\begin{abstract}
Loss functions are error metrics that quantify the difference between a prediction and its corresponding ground truth. Fundamentally, they define a functional landscape for traversal by gradient descent. Although numerous loss functions have been proposed to date in order to handle various machine learning problems, little attention has been given to enhancing these functions to better traverse the loss landscape. In this paper, we simultaneously and significantly mitigate two prominent problems in medical image segmentation namely: i) class imbalance between foreground and background pixels and ii) poor loss function convergence. 
To this end, we propose an Adaptive Logarithmic Loss (ALL) function.  We compare this loss function with the existing state-of-the-art on the ISIC 2018 dataset, the nuclei segmentation dataset as well as the DRIVE retinal vessel segmentation dataset. We measure the performance of our methodology on benchmark metrics and demonstrate state-of-the-art performance. More generally, we show that our system can be used as a framework for better training of deep neural networks.

\keywords{Semantic segmentation  \and class imbalance \and loss functions \and U-Net \and FocusNet.}
\end{abstract}
\section{Introduction}
\label{sec:intro}
With advances in technology, deep convolutional networks have become a fast and accurate means to carry out semantic segmentation tasks. They are widely used in most applications in 2D and 3D medical image analysis. The networks effectively learn to label a binary mask as 0, for every background pixel and as 1 for the foreground. Historically, the Binary Cross Entropy loss emerged as the loss function of choice for this per-pixel labelling task. 
It generally works well for classification and segmentation tasks, as long as the labels for all classes are balanced. If one class dominates over the other, the imbalance results in the network predicting all outputs to be the dominant class, due to convergence to a non optimal local minimum. Some recently proposed loss functions such as the dice loss and the Focal Loss \cite{focal} tackle this problem by weighting some outputs more than others. Other losses such as the Generalized Cross Entropy Loss \cite{genralce} have been shown to be robust to noisy labels.
\begin{figure}[t]
\begin{center}
\includegraphics[width=0.7\linewidth, scale=1.0]{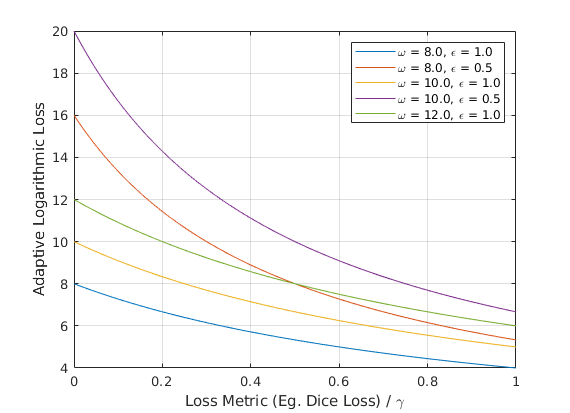}
\end{center}
\caption{The plot shows the value of the derivative of our loss against the value of the Dice Loss that it optimizes. It can be seen that for smaller values of the loss metric, a larger loss is backpropagated. $\gamma$ is fixed empirically based on initial experiments to any value on the $x$-axis.}
\label{derivativeofloss}
\end{figure}
General evaluations of these losses is done by calculating the overall overlap between the ground truth and the prediction. The most basic form of such a metric is the Jaccard Index. 
In contrast, the Dice Index assigns a higher weight to the true positives, and is given by the formula: $DI = \frac{2|G \cap P|}{|G| + |P|}$ where G is the ground truth mask and P is the predicted mask. Due to its high weight on the true positives, $DI$ is also widely used as a loss function. The Tversky Index \cite{tversky} is another proposed metric, that adds further weight to the false positives and false negatives to get better predictions. These similarity metrics are generally converted to loss functions by optimizing over a sum of their class-wise difference from the optimal value. Their general form is $L = \sum_{c} (1 - M)$ where the metric, $M$, can be the Jaccard, Dice or the Tversky Index. The subscript indicates a summation over the number of classes, $c$. Although many loss functions have also been proposed \cite{ce&dice} \cite{dice&focal} \cite{comboloss} as weighted combinations of these losses, none of the existing losses in medical image segmentation explicitly account for both class imbalance, as well as network convergence, even though methods to tackle such problems exist in other computer vision applications \cite{wingloss}.
In this paper, we propose to enhance the properties of the Dice Loss using our methodology. We conduct an extensive hyperparameter search for our loss and empirically show that our technique leads to better convergence of the dice loss under even less optimal settings of the hyperparameters. We compare with state-of-the-art for the same problems, and show performance gains over them. We use the U-Net \cite{unet}, and FocusNet \cite{FocusNet} architectures to compare results. Our enhancement experiments with the Dice Loss due to its popularity in medical image segmentation tasks, but in theory, any loss function could be used here.
The rest of the paper is organized as follows. In Section \ref{sec:loss}, we discuss our loss function. Section \ref{sec:eval} describes our evaluation. Results are presented in Section \ref{sec:results} and we conclude with Section \ref{sec:conc}.

\section{Adaptive Logarithmic Loss}
\label{sec:loss}
We motivate the need for our loss based on the properties a good loss function should possess. Once a loss function computes the error between the label and the ground truth values, the error is backpropagated though a network in order to make it learn. This fundamental task is generally conducted well by all loss functions, though some tend to converge faster than the others. Empirically, Tversky Loss converges in lesser epochs compared to the earlier proposed losses such as CE or the Jaccard Loss. A good loss function should not take too long to converge. It is an added bonus if it speeds up convergence. Secondly, a loss function should be able to adapt to the loss landscape closer to convergence. Keeping these points in mind, we construct a loss function that can both, converge at a faster rate, as well as adaptively refine its landscape when closer to convergence. The formula for this adaptive loss is given by, 
\begin{equation}
	ALL(x)=
	\left\{ \begin{array}{ll}
	 	\omega \ln(1 + \frac{|DL|}{\epsilon}) &|DL| < \gamma \\
	 	|DL|-C &\texttt{otherwise}
	 \end{array} \right.
\end{equation}
where $C = \gamma - \omega \ln(1+ (\frac{\gamma}{\epsilon})$ is used to make the loss function differentiable and smooth at $|DL| = \gamma$ and DL is the computed dice loss. $\gamma, \omega \ and \ \epsilon$ are hypermarameters of this loss function. Further, as the dice loss lies between [0, 1], we experiment with values of $\gamma$ that are [0, 1] to find the optimal threshold to shift to a smoother log based loss for convergence close to the minima. As log is a monotonic function, it smoothens the convergence. The derivative of this loss can be computed via the chain rule. It is visually shown in Figure \ref{derivativeofloss}. Differentiating a function of a function results in the product of two derivatives. This results in, $[ALL(DL(\cdot))]' = ALL'(DL) \ \times \ DL'(\cdot)$, where the plot of $ALL'(DL)$ is shown in Figure \ref{derivativeofloss}. Hence, given any loss as input to our adaptive function, it's derivative will be multiplied by a smooth differentiable function that would in turn remove any discontinuities. 
After experimentation, we observed negligible computational overhead compared to other loss functions for computing our loss. We found the optimal values for the hyperparameters to be, $\gamma = 0.1, \ \omega = 10.0 \ and \ \epsilon = 0.5$. The loss is mainly sensitive to the value of $\gamma$, while $\omega \ and \ \epsilon$ can be kept constant (as we observed little change in the value of the loss across these two parameters once they had been optimally set).
\begin{table}
  \centering
    \begin{tabular}{l|*{6}r}
    \toprule
    \diagbox{$\epsilon$}{$\omega$} & 6 & 8 & 10 & 12 & 14 & 16 \\
    \midrule
    0.3 & 81.43 & 81.48 & 81.51 & 81.59 & 81.90 & 81.67 \\
    0.5 & 81.97 & 81.57 & \textbf{82.43} & 82.24 & 81.58 & 81.07 \\
    1.0 & 81.78 & 82.11 & 81.84 & 81.73 & 82.21 & 81.96 \\
    2.0 & 81.75 & 81.99 & 82.18 & 81.58 & 81.71 & 81.63 \\
    \bottomrule
    \end{tabular}
    \vspace{0.2cm}
  \caption{Optimizing the values of $\omega$ and $\epsilon$ over the corresponding Jaccard Index (\%). Values are averages of 3 runs. Experiments conducted with constant $\gamma=0.1$. JI with baseline Dice Loss = 71.36. Results obtained using FocusNet.}
  \label{tab:exploreomegaepsilon}
  \vspace{-1.0cm}
\end{table}
\begin{table}
\centering
 \begin{tabular}{c| c c c c c c} 
 \hline
$\gamma$ & 0.08 & 0.10 & 0.12 & 0.15 & 0.20 & 0.30\\ 
 \hline
 JI & 81.60 & \textbf{82.43} & 81.54 & 81.51 & 80.85 & 80.97 \\ 
 \hline
 \end{tabular}
  \vspace{0.2cm}
 \caption{Optimizing the values of $\gamma$ over the Jaccard Index (\%). Values are averages of 3 runs. Experiments conducted with constant values of $\omega=10$, $\epsilon=0.5$. JI with baseline Dice Loss = 71.36. Results obtained using FocusNet.}
 \label{tab:varygamma}
\end{table}
\vspace{-.9cm}
\section{Evaluation}
\label{sec:eval}
The experiments for our methodology are conducted with two architectures. We use the benchmark U-Net \cite{unet} and the attention based FocusNet \cite{FocusNet}. A generic U-Net is enhanced with batch normalization, dropout and strided downsampling to improve on it's performance. The FocusNet architecture used is exactly the same as proposed in \cite{FocusNet}. We use 3 datasets that exhibit varying class imbalance for our experiments, to study the effect of our loss on them. The ISIC 2018 skin cancer segmentation dataset \cite{isic2018data}, the data science bowl 2018 cell nuclei segmentation dataset, and the DRIVE retinal vessel segmentation dataset \cite{drivedataset} are used. We do not apply any pre-processing excepting resizing the images to a constant size and scaling the pixel values between $[0, 1]$. For the DRIVE dataset, we extract 200,000 small patches from the images (mostly within the field of view, along with some edge cases) to construct our dataset. The images for the ISIC 2018 dataset were resized to $192 \times 256$, keeping with the aspect ratio of the training set. The images for the cell nuclei segmentation task were resized to $128 \times128$. The patches extracted from the DRIVE dataset were of the size $48 \times 48$. The data for all experiments is divided into a 80:20 split. To keep the evaluation fair, we do not use any augmentation strategies as different augmentations can effect performances differently.  
We apply a grid search style strategy to find the optimal hyperparameters of our loss, where we first run some initial tests to see the behaviour of the loss given some hyperparameters, and then tune them. Initially, we set $\gamma = 0.1$ and tuned the values of $\omega$ and $\epsilon$ to their optimal settings. Then, we use the empirically estimated $\omega$ and $\epsilon$ to find the optimal value for $\gamma$. The values obtained are shown in Table \ref{tab:exploreomegaepsilon} and \ref{tab:varygamma}. From the tables we can see that the loss is not affected a lot by changes in $\omega$ or $\epsilon$, but even small changes in $\gamma$ can cause significant changes in the loss value. This is graphically verified via the derivative of the loss in Figure \ref{derivativeofloss} where we can see that for small values of $\gamma$, the penalty for getting a prediction wrong is a lot larger.
All experiments were run using Keras with a TensorFlow backend. Adam with a learning rate of 1e-4 was used. A constant batch size of 16 was used throughout. Our implementation of the loss will be available on GitHub. The experiments were run for a maximum of 50 epochs.
\begin{figure}[!tbp]
  \begin{subfigure}[b]{0.5\textwidth}
    \includegraphics[width=\textwidth]{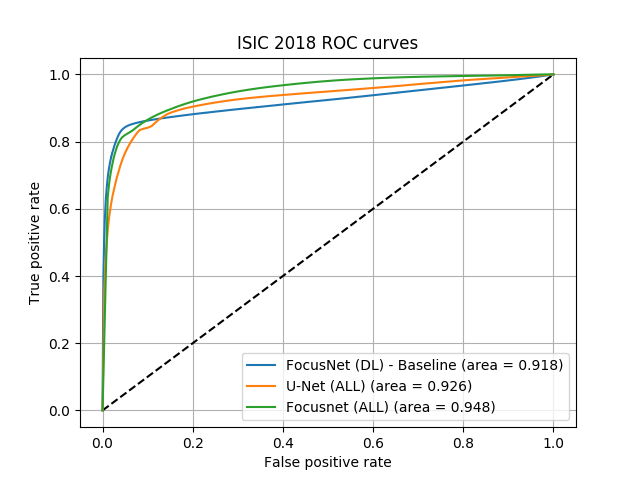}
    \caption{Skin cancer segmentation ROC.}
    \label{isicroc}
  \end{subfigure}
  \hfill
  \begin{subfigure}[b]{0.5\textwidth}
    \includegraphics[width=\textwidth]{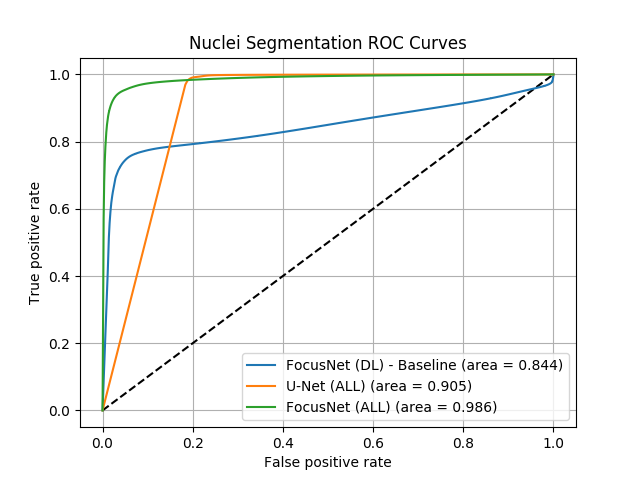}
    \caption{Cell nuclei segmentation ROC.}
    \label{nucleisegroc}
  \end{subfigure}
  \caption{Our loss has a better Area Under the ROC curve than the baseline in both cases. When the imbalance is higher (Figure \ref{nucleisegroc}), our loss provides a much more robust and significant Area Under the ROC curve than the baseline, demonstrating a superior convergence. The curves are plotted for the best performing models for our experiments on the tasks.}
  \vspace{-0.3cm}
\end{figure}
To evaluate the performance of our loss, we compute the intersection over union (IoU) overlap, recall, specificity, F-measure and the area under the receiver operator characteristics curve (AUC-ROC) of the corresponding network predictions trained on various loss functions.

\section{Results}
\label{sec:results}
\begin{table*}[htb]
\label{results_tablemain}
\begin{center}
\begin{tabular}{l||c|c|c||c|c|c||c|c|c}
\hline
\multirow{2}{*}{Method} & \multicolumn{3}{c||}{ISIC $2018$} & \multicolumn{3}{c||}{Data Science Bowl $2018$} & \multicolumn{3}{c}{DRIVE} \\
\cline{2-10}
 & Recall & TNR & Jaccard & Recall & TNR & Jaccard & F1 & Recall & Jaccard \\
\hline
\hline
\cite{unet} (JI) & 78.62 & 85.21 & 72.96 & 76.27 & 81.29 & 73.64 & 78.46 & 73.28 & 65.37 \\
\hline
\cite{unet} (DL) & 76.12 & 83.74 & 69.34 & 74.92 & 82.85 & 64.57 & 78.94 & 74.10 & 67.79 \\
\hline
\cite{unet} (TL) & 80.82 & 86.98 & 74.18 & 79.21 & 85.81 & 77.72 & 79.89 & 74.47 & 66.18 \\
\hline
\cite{unet} (FL) & \textbf{83.76} & \textbf{89.85} & \textbf{79.17} & 78.27 & 86.88 & 78.82 & 80.07 & 75.65 & 68.96 \\
\hline
\cite{unet} (CL) & 82.19 & 87.96 & 75.87 & 77.34 & 85.63 & 78.24 & 79.26 & 74.33 & 67.42 \\
\hline
\cite{unet} (\textbf{ALL}) & 83.56 & 88.47 & 77.69 & \textbf{79.88} & \textbf{87.27} & \textbf{79.71} & \textbf{81.41} & \textbf{75.83} & \textbf{69.23} \\
\hline
\hline
\cite{FocusNet} (JI) & 80.13 & 86.17 & 72.28 & 77.12 & 84.19 & 74.97 & 77.87 & 73.28 & 64.67 \\
\hline
\cite{FocusNet} (DL) & 80.78 & 85.81 & 71.92 & 78.37 & 84.92 & 77.82 & 77.86 & 73.98 & 64.71 \\
\hline
\cite{FocusNet} (TL) & 84.86 & 90.62 & 77.63 & 79.64 & 88.27 & 77.28 & 81.31 & 74.19 & 68.66 \\
\hline
\cite{FocusNet} (FL) & 86.19 & \textbf{93.95} & 82.78 & 79.26 & 89.17 & 78.73 & 81.28 & \textbf{76.89} & 69.57 \\
\hline
\cite{FocusNet} (CL) & 84.82 & 86,19 & 78.63 & 80.65 & 87.34 & 79.35 & 78.64 & 74.18 & 68.34 \\
\hline
\cite{FocusNet} (\textbf{ALL}) & \textbf{86.62} & 92.78 & \textbf{82.84} & \textbf{82.51} & \textbf{90.86} & \textbf{81.37} & \textbf{82.17} & 76.13 & \textbf{70.96} \\
\hline
\end{tabular}
\end{center}
\caption{Segmentation results for the three datasets. All values in the ISIC 2018 experiments, Data Science Bowl and the DRIVE retinal blood vessel segmentation datasets are averaged over 5, 3 and 2 runs respectively to average out the effects of random weight initialization as much as possible. The values reported are all in \%. Here, \cite{unet} and \cite{FocusNet} are the U-Net and the FocusNet architectures respectively, with the relevant loss function. TNR is the True Negative Rate (Specificity).}
\vspace{-0.2cm}
\end{table*}
\begin{table}
\centering
 \begin{tabular}{c|| c| c| c} 
 \hline
 Method & Dice & Precision & Recall \\ 
 \hline
U-Net (FTL)  & 82.92 & 79.74 & 92.61 \\ 
 \hline
Att-U-Net+M+D (FTL)  & 85.61 & 85.82 & 89.71 \\ 
 \hline
FocusNet (\textbf{ALL})  & \textbf{87.14} & \textbf{88.11} & \textbf{90.47} \\ 
 \hline
 \end{tabular}
 \vspace{0.2cm}
 \caption{Experiments run for the ISIC 2018 dataset training-validation-test split in \cite{ftl}. Our reported values (in \%) are averaged over 3 runs. `M' denotes Multi Scale Input. `D' denotes deep supervision.}
 \label{tab:isic2018ftl}
 \vspace{-0.5cm}
\end{table}
\begin{figure}[htb]
\begin{center}
\includegraphics[width=0.7\linewidth, scale=1.0]{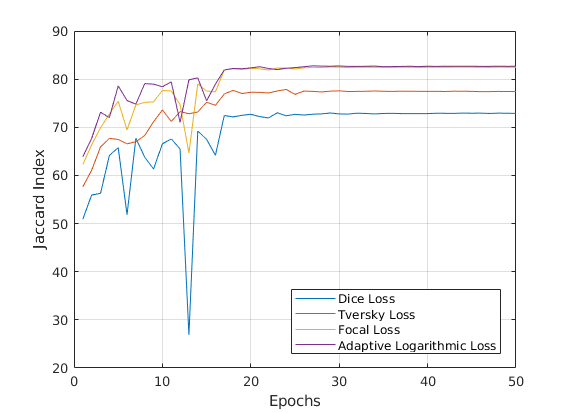}
\end{center}
\caption{Comparing the influence of different loss functions on FocusNet. The plot shows the validation Jaccard Index on the ISIC 2018 dataset vs the number of epochs.}
\label{lossconvergenceplotFocusNet}
\end{figure}
We compare the performance of our loss (Table 3) with the Jaccard Loss (JL), Dice Loss (DL), Tversky Loss (TL) \cite{tversky}, Focal Loss (FL) \cite{focal} and the Combo Loss (CL) \cite{comboloss}. The ISIC 2018 dataset shows the least imbalance and hence the results are fairly even for this dataset, though our loss does manage to get the best in class IoU. This is also exhibited in Figure \ref{isicroc} where we plot the ROC curves. FocusNet with the ALL gets the best area under the curve. We also compared FocusNet trained with the ALL against the recently proposed Focal Tversky Loss \cite{ftl} for the ISIC 2018 dataset. We used the same train-test split as their implementation based on their open sourced GitHub code and averaged our results over 3 runs. Our loss outperforms their architecture trained on their loss by 1.53 \% on the Dice Index. We also report a better precision and recall than their methodology. The results for this experiment are shown in Table \ref{tab:isic2018ftl}. The nuclei segmentation dataset exhibits more class imbalance, and in such a case our loss shows significantly superior performance compared to all the other losses. We get significant gains over the baseline AUC as shown in Figure \ref{nucleisegroc}. FocusNet with just the Dice Loss suffers from poor convergence and does not get competitive results. We do not compare the DRIVE dataset by the AUC or accuracy as these metrics are fairly saturated for this dataset and do not offer any statistically significant insights. It is interesting to note that we do get an improved F-measure score and the best in class IoU, which given the large number of patches extracted, is statistically significant. Overall, our loss shows significantly better performance than the baseline Dice Loss for all three datasets, which means that it manages to optimize the loss to a significantly better minimum on the loss landscape leading to a more optimal solution. In all cases, the trend shown by our loss is to converge to within delta of the optimal solution and then refine the convergences using the adaptive strategy. Without the adaptive strategy, our loss often gets stuck in local minimum, which reiterates the importance of having such a piece-wise continuous loss. The other loss functions (especially Dice Loss) exhibit slightly unstable convergence. We observed that our loss mostly converged faster than JL, DL, TL and CL. FL convergence is at par with our loss, while TL converges more smoothly. We verify this visually by plotting the behaviour of the losses against the number of epochs (see Figure \ref{lossconvergenceplotFocusNet}). Compared to the other losses, we observed no extra computational overhead and the need to tune just one hyperparameter during training. This coupled with a better handling of class imbalance makes our loss a superior choice for such class of problems.

\section{Conclusion}
\label{sec:conc}
In this paper, we proposed an enhanced loss function that potentially constructs a loss landscape that is easier to traverse via backpropagation. We tested our approach on 3 datasets that show varied class imbalance. As the imbalance in the data increases, our loss provides a more robust solution (along with better convergence) to this prominent problem compared to other state-of-the-art loss functions. We base our conclusions on carefully constructed evaluation metrics for each task that show significantly superior performance in favour of our loss compared to the baseline.

\bibliographystyle{splncs04}
\bibliography{refs}

\end{document}